\documentclass[aps,prl,preprint,tightenlines,superscriptaddress,showpacs,byrevtex]{revtex4}
\usepackage{amsmath}
\usepackage{epsfig}
\begin{document}

\epsfysize3cm
\begin{flushright}

\hspace*{4.5in}BELLE-CONF-0747\\

\end{flushright}                   


\title{\quad\\[0.5cm] \Large
Search for $B^0 \to \rho^0 \rho^0$ and Non-Resonant 
$B^0 \to 4\pi$ Decays}
\tighten
\affiliation{Budker Institute of Nuclear Physics, Novosibirsk}
\affiliation{Chiba University, Chiba}
\affiliation{University of Cincinnati, Cincinnati, Ohio 45221}
\affiliation{Department of Physics, Fu Jen Catholic University, Taipei}
\affiliation{Justus-Liebig-Universit\"at Gie\ss{}en, Gie\ss{}en}
\affiliation{The Graduate University for Advanced Studies, Hayama}
\affiliation{Gyeongsang National University, Chinju}
\affiliation{Hanyang University, Seoul}
\affiliation{University of Hawaii, Honolulu, Hawaii 96822}
\affiliation{High Energy Accelerator Research Organization (KEK), Tsukuba}
\affiliation{Hiroshima Institute of Technology, Hiroshima}
\affiliation{University of Illinois at Urbana-Champaign, Urbana, Illinois 61801}
\affiliation{Institute of High Energy Physics, Chinese Academy of Sciences, Beijing}
\affiliation{Institute of High Energy Physics, Vienna}
\affiliation{Institute of High Energy Physics, Protvino}
\affiliation{Institute for Theoretical and Experimental Physics, Moscow}
\affiliation{J. Stefan Institute, Ljubljana}
\affiliation{Kanagawa University, Yokohama}
\affiliation{Korea University, Seoul}
\affiliation{Kyoto University, Kyoto}
\affiliation{Kyungpook National University, Taegu}
\affiliation{Ecole Polyt\'ecnique F\'ed\'erale Lausanne, EPFL, Lausanne}
\affiliation{University of Ljubljana, Ljubljana}
\affiliation{University of Maribor, Maribor}
\affiliation{University of Melbourne, School of Physics, Victoria 3010}
\affiliation{Nagoya University, Nagoya}
\affiliation{Nara Women's University, Nara}
\affiliation{National Central University, Chung-li}
\affiliation{National United University, Miao Li}
\affiliation{Department of Physics, National Taiwan University, Taipei}
\affiliation{H. Niewodniczanski Institute of Nuclear Physics, Krakow}
\affiliation{Nippon Dental University, Niigata}
\affiliation{Niigata University, Niigata}
\affiliation{University of Nova Gorica, Nova Gorica}
\affiliation{Osaka City University, Osaka}
\affiliation{Osaka University, Osaka}
\affiliation{Panjab University, Chandigarh}
\affiliation{Peking University, Beijing}
\affiliation{University of Pittsburgh, Pittsburgh, Pennsylvania 15260}
\affiliation{Princeton University, Princeton, New Jersey 08544}
\affiliation{RIKEN BNL Research Center, Upton, New York 11973}
\affiliation{Saga University, Saga}
\affiliation{University of Science and Technology of China, Hefei}
\affiliation{Seoul National University, Seoul}
\affiliation{Shinshu University, Nagano}
\affiliation{Sungkyunkwan University, Suwon}
\affiliation{University of Sydney, Sydney, New South Wales}
\affiliation{Tata Institute of Fundamental Research, Mumbai}
\affiliation{Toho University, Funabashi}
\affiliation{Tohoku Gakuin University, Tagajo}
\affiliation{Tohoku University, Sendai}
\affiliation{Department of Physics, University of Tokyo, Tokyo}
\affiliation{Tokyo Institute of Technology, Tokyo}
\affiliation{Tokyo Metropolitan University, Tokyo}
\affiliation{Tokyo University of Agriculture and Technology, Tokyo}
\affiliation{Toyama National College of Maritime Technology, Toyama}
\affiliation{Virginia Polytechnic Institute and State University, Blacksburg, Virginia 24061}
\affiliation{Yonsei University, Seoul}
  \author{K.~Abe}\affiliation{High Energy Accelerator Research Organization (KEK), Tsukuba} 
  \author{I.~Adachi}\affiliation{High Energy Accelerator Research Organization (KEK), Tsukuba} 
  \author{H.~Aihara}\affiliation{Department of Physics, University of Tokyo, Tokyo} 
  \author{K.~Arinstein}\affiliation{Budker Institute of Nuclear Physics, Novosibirsk} 
  \author{T.~Aso}\affiliation{Toyama National College of Maritime Technology, Toyama} 
  \author{V.~Aulchenko}\affiliation{Budker Institute of Nuclear Physics, Novosibirsk} 
  \author{T.~Aushev}\affiliation{Ecole Polyt\'ecnique F\'ed\'erale Lausanne, EPFL, Lausanne}\affiliation{Institute for Theoretical and Experimental Physics, Moscow} 
  \author{T.~Aziz}\affiliation{Tata Institute of Fundamental Research, Mumbai} 
  \author{S.~Bahinipati}\affiliation{University of Cincinnati, Cincinnati, Ohio 45221} 
  \author{A.~M.~Bakich}\affiliation{University of Sydney, Sydney, New South Wales} 
% ...
\collaboration{The Belle Collaboration}

\tighten


\begin{abstract}
We search for the decay $B^0\to\rho^0\rho^0$ and other possible charmless modes 
with a $\pi^+\pi^-\pi^+\pi^-$ final state, 
including $B^0\to\rho^0f_0(980)$, $B^0\to f_0(980)f_0(980)$, 
$B^0\to f_0(980)\pi\pi$, $B^0\to\rho^0\pi\pi$ and non-resonant $B^0\to 4\pi$. 
These results are obtained from a data sample containing $520 \times 10^6$ 
$B \overline B$ pairs collected by the Belle detector at the KEKB 
asymmetric-energy $e^+e^-$ collider. 
We measure a branching fraction of $(0.9\pm0.4^{+0.3}_{-0.4})\times 10^{-6}$, 
or $\mathcal{B}(B^0\to\rho^0\rho^0)<1.6\times 10^{-6}$ at the 90\% 
confidence level. The significance including systematic uncertainties is
1.8$\sigma$. These values correspond to the final state being longitudinally 
polarized. 
We also measure the branching fraction of non-resonant $B^0\to 4\pi$ decay 
to be $(10.2\pm4.7^{+2.3}_{-1.5})\times 10^{-6}$ with 2.1$\sigma$ 
significance, and set the 90\% confidence level upper limit 
$\mathcal{B}(B^0\to 4\pi)<17.3\times 10^{-6}$. 
For the other related decays, 
$B^0\to\rho^0f_0(980)$, $B^0\to f_0(980)f_0(980)$, $B^0\to f_0(980)\pi\pi$ 
and $B^0\to\rho^0\pi\pi$, no significant signals are observed and upper 
limits on the branching fractions are set.
\end{abstract}


\maketitle

\tighten

\normalsize

\begin{center}
\begin{Large}
{\bf INTRODUCTION\\}
\end{Large}
\end{center}

In the Standard Model (SM), $CP$ violation in the weak interaction can be 
measured through the differences between $B$ and $\overline B$ mesons decays. 
Measurements of $CP$ violation help determine (or constrain) the elements of 
the Cabibbo-Kobayashi-Maskawa (CKM) quark-mixing matrix \cite{1} and thus 
test the Standard Model. 
The time-dependent $CP$ asymmetry for the decay of a neutral $B$ meson via a 
$b\to u$ process into a $CP$ eigenstate can determine the CKM phase angel 
$\phi_2 \equiv - \frac{V_{td}V_{tb}^*}{V_{ud}V_{ub}^*}$. However, in addition 
to the $b \to u$ tree amplitude, a $b \to d$ penguin amplitude contributions, 
and thus an isospin analysis \cite{2} is needed to determine $\phi_2$.

At present, the $\phi_2$ constraints on $B\to\pi\pi$ \cite{201}, 
$B\to\rho\pi$ \cite{202} and $B\to\rho\rho$ \cite{203} are well studied. 
However, it is necessary to measure the branching fraction and polarization of 
$B^0\to\rho^0\rho^0$ for improved $\phi_2$ constraints. 
Angular analysis can provide additional information on $VV$ decays such as 
$B\to\rho\rho$.
Polarization measurements in the $B^0\to\rho^+ \rho^-$ and 
$B^+\to\rho^+ \rho^0$ modes \cite{203} show the dominance of longitudinal 
polarization, thus $B^0\to \rho^+ \rho^-$ is a $CP$ eigenstate; measurements of
the branching fraction and polarization of $CP$-violating asymmetry in 
$B^0\to\rho^0\rho^0$ decays would complete the isospin triangle and improve 
the constraints on $\phi_2$.

Theoretically, the tree contribution to $B^0\to \rho^0\rho^0$ is 
color-suppressed and thus its branching fraction is much smaller than that of 
$B^0\to\rho^+ \rho^-$ or $B^+\to\rho^+ \rho^0$.  The decay rate for 
$\rho^0 \rho^0$ is sensitive to the penguin amplitude. Predictions for 
$B^0\to\rho^0\rho^0$ using perturbative QCD (pQCD) \cite{32} or QCD 
factorization \cite{13, 14} approaches suggest that the branching fraction 
$\mathcal{B}(B^0\to\rho^0\rho^0)$ is at or below $1 \times 10^{-6}$ and 
that the longitudinal polarization fraction, $f_L$, is around 0.85. A non-zero 
branching fraction for $B^0\to\rho^0\rho^0$ was first reported by the BaBar 
collaboration \cite{11}; they measured a branching fraction of 
$\mathcal{B}(B^0\to\rho^0\rho^0)=(1.07\pm 0.33\pm 0.19)\times 10^{-6}$ with 
a significance of 3.5 standard deviations ($\sigma$), and a longitudinal 
polarization fraction, $f_L=0.87\pm 0.13 \pm 0.04$. 

A theoretical prediction for the non-resonant $B^0\to 4\pi$ branching fraction 
is around $1 \times 10^{-4}$ \cite{34}. The most recent measurement of this 
decay was made by the DELPHI collaboration \cite{33}, who set a 90\% 
confidence level upper limit on the branching fraction of 
$2.3 \times 10^{-4}$.

In this paper, we report the results of a search for 
$B^0\to\rho^0\rho^0$ along with other modes, including 
$B^0\to \rho^0f_0$, $B^0\to f_0f_0$, $B^0\to f_0\pi\pi$, $B^0\to \rho^0\pi\pi$ 
and non-resonant $B^0\to 4\pi$.

\begin{center}
\begin{Large}
{\bf DATA SET AND APPARATUS\\}
\end{Large}
\end{center}

The data sample used contains $520 \times 10^6 \ B \overline B$ pairs collected 
with the Belle detector at the KEKB asymmetric-energy $e^+e^-$ (3.5 and 8 GeV) 
collider \cite{101}, operating at the $\Upsilon(4S)$ resonance. The Belle 
detector \cite{102, 103} is a large-solid-angle magnetic spectrometer that 
consists of a silicon vertex detector, a 50-layer central drift chamber (CDC), 
an array of aerogel threshold Cherenkov counters (ACC), a barrel-like 
arrangement of time-of-flight scintillation counters (TOF), and an 
electromagnetic calorimeter comprised of CsI(Tl) crystals (ECL) located inside 
a superconducting solenoid coil that provides a 1.5 T magnetic field. An iron 
flux-return located outside of the coil is instrumented to detect $K^{0}_{L}$ 
mesons and to identify muons.

We study signal and backgrounds using Monte Carlo (MC) simulation. For these 
simulations, signal decays, generic $b\to c$ decays and charmless rare $B$ 
decays are generated with the EVTGEN \cite{109} event generator. Signal MC 
event generation utilizes the PHOTOS simulation package to take account of 
final-state radiation \cite{105}. The continuum MC events are generated 
through $e^+e^- \to \gamma^* \to q\overline q$ $(q=u,d,s,c)$ decays in 
JETSET \cite{110}. The GEANT3 \cite{111} package is used for detector 
simulation.

\begin{center}
\begin{Large}
{\bf EVENT SELECTION AND RECONSTRUCTION \\}
\end{Large}
\end{center}

$B^0$ meson candidates are reconstructed from neutral combinations of four 
charged pions. Charged track candidates are required to have a 
distance-of-closest-approach to the interaction point (IP) of less 
than 2 cm in the beam 
direction ($z$-axis) and less than 0.1 cm in the transverse plane; they are 
also required to have a transverse momentum $p_T>0.1$ GeV/$c$ in the 
laboratory frame.  
Charged pions are identified using particle identification (PID) 
information obtained from the the CDC ($dE/dx$), the ACC and the TOF. We 
distinguish charged kaons and pions using a likelihood ratio 
$\mathcal{R}_{\mathrm{PID}}=\mathcal{L}_{K}/(\mathcal{L}_{K}+
\mathcal{L}_{\pi})$, where $\mathcal{L}_{\pi}(\mathcal{L}_{K})$ is a 
likelihood value for the pion (kaon) hypothesis. We require 
$\mathcal{R}_{\mathrm{PID}} < 0.4$ for the four charged pions. The pion
identification efficiency is 90\%, and 12\% of kaons are 
misidentified as pions. Charged particles positively identified as an 
electron or a muon are removed.

To veto $B\to D^{(*)}\pi$ backgrounds, we remove candidates 
that satisfy any one of the following conditions: 
$|M(h^{\pm}\pi^{\mp}\pi^{\mp})-M_{D_{(s)}}|<13 \ \mathrm{MeV}/c^2$ or 
$|M(h^{\pm}\pi^{\mp})-M_{D^0}|<13 \ \mathrm{MeV}/c^2$, where $h^{\pm}$ is 
either a pion or a kaon, and $M_{D_{(s)}}$ and $M_{D^0}$ are the nominal 
masses of $D_{(s)}$ and $D^0$, respectively. Furthermore, to reduce the 
$B^0\to a_1^\pm \pi^\mp$ feeddown in the signal region 
($0.55\ \mathrm{GeV}/c^2<M_{1,2}(\pi\pi)<1.35\ \mathrm{GeV}/c^2$), we 
require that the highest momentum pion have a momentum in the $\Upsilon(4S)$ 
center-of-mass (CM) frame within the range 1.30-2.65 GeV/$c$.

The signal event candidates are characterized by two kinematic variables: 
beam-energy constrained mass, $M_{\mathrm{bc}}=\sqrt{E^2_{\mathrm{beam}}-
P^2_{B}}$, and energy difference, $\Delta E = E_{B}-E_{\mathrm{beam}}$, 
where $E_{\mathrm{beam}}$ is the run-dependent beam energy, $P_B$ and $E_B$ 
are the momentum and energy of the $B$ candidate in the $\Upsilon(4S)$ 
CM frame. We select candidate events in the region 
$5.24\ \mathrm{GeV}/c^2 < M_{\mathrm{bc}} < 5.30\ \mathrm{GeV}/c^2$ and 
$|\Delta E| < 0.01$ GeV.

The invariant masses $M_1(\pi^+\pi^-)$ and $M_2(\pi^+\pi^-)$, are used to 
distinguish different modes from signal. In $B^0 \to \rho^0\rho^0 \to 
(\pi^+\pi^-)(\pi^+\pi^-)$ decays, there are two possible combinations: 
$(\pi^+_1\pi^-_1)(\pi^+_2\pi^-_2)$ and $(\pi^+_1\pi^-_2)(\pi^+_2\pi^-_1)$, 
where the subscripts label the momentum ordering (e.g., $\pi^+_1$ has 
higher momentum than $\pi^+_2$).
According to a MC study, 
assuming $B^0\to\rho^0\rho^0$ decays are longitudinally polarized, the 
reconstructed $\pi^+\pi^-$ pair containing a high momentum $\pi^+(\pi^-)$ and 
a low momentum $\pi^-(\pi^+)$ will corresponds to the correct combination 
84.8\% of the time. Here we consider both possible $\pi^+\pi^-$ combinations 
and select candidate events if either one of the combinations lies in the 
$\rho^0\rho^0$ signal mass window, which is 
$0.55\ \mathrm{GeV}/c^2 < M(\pi^+_1\pi^-_1)
\cap M(\pi^+_2\pi^-_2) < 1.35\ \mathrm{GeV}/c^2$ or 
$0.55\ \mathrm{GeV}/c^2 < M(\pi^+_1\pi^-_2) \cap M(\pi^+_2\pi^-_1) 
< 1.35\ \mathrm{GeV}/c^2$. If a candidate event has two $\pi^+\pi^-$ pair 
combinations that both lie in the $\rho^0\rho^0$ signal mass window, we cannot 
distinguish which $\rho^0\rho^0$ mass combination is correct. In such cases, 
we select the $\pi^+\pi^-$ pair containing a high momentum $\pi^+(\pi^-)$ and 
a low momentum $\pi^-(\pi^+)$ as the correct combination; 
with this selection, 0.2\% of the signal is incorrectly reconstructed 
according to the MC. 
For fitting, we randomly assign the $\pi^+\pi^-$ pairs to 
either $M_1(\pi^+\pi^-)$ or $M_2(\pi^+\pi^-)$ to symmetrize the 2-D invariant 
mass distribution. Therefore, the probability density functions (PDF) for 
$M_1(\pi^+\pi^-)$-$M_2(\pi^+\pi^-)$ are symmetric in both the 
$M_1(\pi^+\pi^-)$ and $M_2(\pi^+\pi^-)$ projections.

\begin{center}
\begin{Large}
{\bf BACKGROUND SUPPRESSION \\}
\end{Large}
\end{center}

The dominant background is the continuum. 
To distinguish signal from the jet-like continuum background, 
we use modified Fox-Wolfram moments \cite{104}, which are combined 
into a Fisher discriminant. 
This discriminant is combined with PDFs for the cosine of the $B$ 
flight direction in the CM and the distance 
in the $z$-direction between two $B$ mesons to form a likelihood ratio 
$\mathcal{R}=\mathcal{L}_{s}/(\mathcal{L}_{s}+\mathcal{L}_{q\overline q})$. 
Here, $\mathcal{L}_{s}$ ($\mathcal{L}_{q\overline q}$) is a likelihood
function for signal (continuum) events that is obtained
from the signal MC simulation (events in the sideband region 
$M_{\rm bc}<5.26$~GeV/c$^2$).
We also use the flavor tagging quality variable $r$ 
provided by a tagging algorithm~\cite{112} that identifies
the flavor of the accompanying $B^0$ meson in the
$\Upsilon(4S)\to B^0\overline B^0$.
The variable $r$ ranges from $r=0$ (no flavor discrimination)
to $r=1$ (unambiguous flavor assignment), and
is used to divide the data sample into six $r$ bins.
Since the discrimination between signal and continuum events
depends on the $r$-bin, we impose different requirements
on $\mathcal{R}$ for each $r$-bin.
We determine the $\mathcal{R}$ requirement so that it
maximizes the figure-of-merit $N_s / \sqrt{N_s + N_{q\overline q}}$,
where $N_s$ $(N_{q\overline q})$ is the expected number of
signal (continuum) events in the signal region 
($|\Delta E|<0.05$ GeV and $5.27 \ \mathrm{GeV}/c^2 < M_{\mathrm{bc}} 
< 5.29 \ \mathrm{GeV}/c^2$).

After applying all selection criteria, 17\% of selected events have multiple 
$B^0\to\rho^0\rho^0$ candidates. For these events we select a single 
candidate that having the smallest $\chi^2$ value of the $B^0$ decay 
vertex reconstruction.
The detection efficiency for the signal MC 
is calculated to be 7.11\% (9.57\%) 
for longitudinal (transverse) polarization.

\begin{center}
\begin{Large}
{\bf ANALYSIS PROCEDURE \\}
\end{Large}
\end{center}

\begin{figure}[htbp]
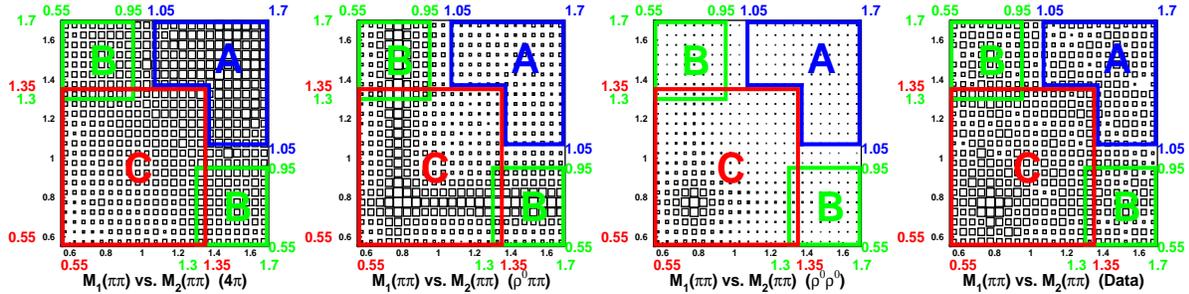

\centering
\epsfig{file=./rho0rho0/fpi-grand.epsi,width=1.5in,height=1.5in}
\epsfig{file=./rho0rho0/rho0pipi-grand.epsi,width=1.5in,height=1.5in}
\epsfig{file=./rho0rho0/sigmc00-grand.epsi,width=1.5in,height=1.5in}
\epsfig{file=./rho0rho0/data.epsi,width=1.5in,height=1.5in}
\caption{$M_1(\pi\pi)$-$M_2(\pi\pi)$ distributions in MC simulation 
of non-resonant $B^0\to 4\pi$, $B^0\to \rho^0\pi\pi$, and 
$B^0\to \rho^0\rho^0$ decays, and the data, from left to right. Non-resonant 
$B^0\to 4\pi$ and $B^0\to \rho^0\pi\pi$ MC decays are generated with 
4- and 3-body phase space models.}
\label{mpipi}
\end{figure}

Since there are large overlaps between $B^0\to\rho^0\rho^0$ and other signal 
decay modes in the $M_1(\pi\pi)$-$M_2(\pi\pi)$ distribution, it is better 
to distinguish these modes using a simultaneous fit to a large 
$M_1(\pi\pi)$-$M_2(\pi\pi)$ region. 
However, the correlations between ($\Delta E$, $M_{\mathrm{bc}}$) and 
($M_1$, $M_2$) for backgrounds and signals will lead to large uncertainties 
as the region in $M_1(\pi\pi)$-$M_2(\pi\pi)$ used in a simultaneous fit 
increases. 
The $M_1(\pi\pi)$-$M_2(\pi\pi)$ distribution is shown in 
Fig.~\ref{mpipi} for Monte Carlo samples of non-resonant $B^0\to 4\pi$, 
$B^0\to \rho^0\pi\pi$, $B^0\to \rho^0\rho^0$ decays, and the data. 
The analysis proceeds in three setps: 
we first measure the non-resonant $B^0\to 4\pi$ 
branching fraction in area A, then measure the $B^0\to \rho^0\pi\pi$ 
in area B by fixing the branching fraction of non-resonant $B^0\to 4\pi$
determined in area A. 
Finally, we determine $B^0\to \rho^0\rho^0$ and the other 
possible decay modes in area C 
by fixing the branching fractions of the non-resonant $B^0\to 4\pi$ and 
$B^0\to \rho^0\pi\pi$ decays.
This procedure minimizes bias caused by unknown correlations between variables
($\Delta E$, $M_{\mathrm{bc}}$) and variables ($M_1$, $M_2$). 

For the non-resonant $B^0\to 4\pi$ branching fraction measurement, 
we use the nominal $\pi^+\pi^-$ mass region 
$1.05\ \mathrm{GeV}/c^2<M_{1(2)}(\pi\pi)<1.70\ \mathrm{GeV}/c^2$ and 
$1.35\ \mathrm{GeV}/c^2<M_{2(1)}(\pi\pi)<1.70\ \mathrm{GeV}/c^2$ 
(area A), 
in which only $B^0\to a_1^{\pm}\pi^{\mp}$ and non-resonant $B^0\to 4\pi$ 
signals having the same final state need to be considered. 
For the $B^0\to \rho^0\pi\pi$ branching fraction measurement, 
we use events in the mass regions 
$1.30\ \mathrm{GeV}/c^2<M_{1(2)}(\pi\pi)<1.70\ \mathrm{GeV}/c^2$ and
$0.55\ \mathrm{GeV}/c^2<M_{2(1)}(\pi\pi)<0.95\ \mathrm{GeV}/c^2$
(area B), 
in which $B^0\to a_1^{\pm}\pi^{\mp}$, non-resonant $B^0\to 4\pi$ 
and $B^0\to\rho^0\pi\pi$ decays are included.
The yields for $B^0\to \rho^0\rho^0$, $\rho^0f_0$, $f_0f_0$ and $f_0\pi\pi$ 
decays are measured using a simultaneous fit in which the branching 
fractions of non-resonant $B^0\to 4\pi$ and $B^0\to \rho^0\pi\pi$ 
in the mass region 
$0.55\ \mathrm{GeV}/c^2<M_{1,2}(\pi\pi)<1.35\ \mathrm{GeV}/c^2$
(area C) are fixed.

In all fits, we fix the branching fraction of 
$B^0\to a_1^{\pm}\pi^{\mp}$ to the published value 
$(33.2\pm 3.0\pm 3.8)\times10^{-6}$~\cite{107}. 
Recently, Belle presented a preliminary result 
for the $B^0\to a_1^{\pm}\pi^{\mp}$ branching fraction~\cite{41}, 
which is consistent with BaBar's value.
We also float the $B^0\to a_1^{\pm}\pi^{\mp}$ yield in the fit; the 
result with its error is consistent with the assumed value.

The signal yields are extracted by performing 
extended unbinned maximum likelihood (ML) fits.
In the fits, we use four dimensional 
($M_{\mathrm{bc}}$, $\Delta E$, $M_1$, $M_2$) information 
to measure the branching fraction of $B^0$ decays into
non-resonant $4\pi$, $\rho^0\rho^0$, $\rho^0f_0$, 
$f_0f_0$ and $f_0\pi\pi$, 
and two dimensional ($M_{\mathrm{bc}}$, $\Delta E$) information 
for the decay $B^0\to \rho^0\pi\pi$. 
We perform a 2D fit to extract the $\rho^0\pi\pi$ yield 
since the $M_1(\pi\pi)$-$M_2(\pi\pi)$ region used in the fit
is expected to consist of $a_1^{\pm}\pi^{\mp}$, 
non-resonant $4\pi$ and $\rho^0\pi\pi$ decays and 
no contributions from other resonant decay modes. 
We define the likelihood function 
\begin{equation}
\mathcal{L}= \exp \biggl(-\sum_{j} n_j \biggr)
             \prod^{\mathrm{N_{cand}}}_{i=1}\biggl(\sum_{j} n_j P^i_j \biggr),
\end{equation}
where $i$ is the event identifier, 
$j$ indicates one of the event type categories for
signals and backgrounds;
$n_j$ denotes the yield of the $j$-th category, 
and $P^i_j$ is the probability density function (PDF)
for the $j$-th category. 
For the 4D fits, the PDFs are a product 
of two smoothed two-dimensional functions: 
$P^i_j=P_j(M^i_{\mathrm{bc}}, \Delta E^i, M^i_1, M^i_2)= 
f_{\mathrm{smoothed}}(M^i_{\mathrm{bc}}, \Delta E^i)
\times f_{\mathrm{smoothed}}(M^i_1, M^i_2)$. 
For the $B^0\to \rho^0\pi\pi$ branching fraction measurement, 
the PDFs are two-dimensional functions, i.e. 
$P^i_j=P_j(M^i_{\mathrm{bc}}, \Delta E^i)
=f_{\mathrm{smoothed}}(M^i_{\mathrm{bc}}, \Delta E^i)$. 

For the signal modes,
the smoothed functions
$f_{\mathrm{smoothed}}(M^i_{\mathrm{bc}}, \Delta E^i)$ 
and $f_{\mathrm{smoothed}}(M^i_1, M^i_2)$ are 
obtained from MC simulations.
For the $M_{\mathrm{bc}}$ and $\Delta E$ PDFs,
possible differences between the real data and the MC modeling
are calibrated using a large control sample of
$B^0 \to D^-(K\pi\pi)\pi^+$ decays. 
We find 16.2\% (3.8\%) of the reconstructed signal MC events are
self-cross-feed (SCF) events for longitudinal (transverse) polarization; 
for SCF events at least one track from the decay $B^0\to\rho^0\rho^0$ 
is replaced by one from the accompanying $B$ meson decay.
We use different PDFs for the SCF events and correctly reconstructed
events, and employ the SCF fraction 
for a longitudinally polarized decay in 
the nominal fit.
The uncertainty in the fraction of transversely polarized 
$B^0\to \rho^0\rho^0$ is taken into account as a systematic error.

For the continuum and charm $B$ decay backgrounds,
we use a linear function for $\Delta E$, 
an ARGUS function~\cite{106} for $M_{\mathrm{bc}}$
and a two-dimensional smoothed function 
for $M_1$-$M_2$.
The parameters of the linear function and ARGUS function for
the continuum events are floated in the fit.
Other parameters and shape of the $M_1$-$M_2$ functions are obtained
from MC simulations and fixed in the fit.

For the charmless $B$ decay backgrounds,
we construct four separate PDFs for $B^0\to a_1^{\pm}\pi^{\mp}$,
$B^+\to\rho^+\rho^0$, $B\to\rho^+\rho^-+\rho\pi$ and other
charmless $B$ decays; all the PDFs are obtained using MC simulations.
In the fit, while the yields of the $B^0\to a_1^{\pm}\pi^{\mp}$
, $B^+\to\rho^+\rho^0$ and $B\to\rho^+\rho^-+\rho\pi$ are fixed to 
expected values obtained from measured branching fractions,
the yield of other charmless $B$ decays is floated.

\begin{center}
\begin{Large}
{\bf MEASUREMENTS OF BRANCHING FRACTIONS \\}
\end{Large}
\end{center}

With the fitted signal yields $n_{\mathrm{sig.}}$, 
we calculate the branching fraction $\mathcal{B}$ using
\begin{equation}
\mathcal{B} = {{n_{\mathrm{sig.}}} \over {N_{B \overline B} 
\cdot \epsilon_{\mathrm{MC}} \cdot \epsilon_{\mathrm{PID}}}},
\end{equation}
where $\epsilon_{\mathrm{MC}}$ is 
the overall reconstruction efficiency 
obtained using MC samples, 
$\epsilon_{\mathrm{PID}}$ is a PID efficiency correction
that takes into account the efficiency difference between data and MC,
and $N_{B\overline B}$ is the number of $B\overline B$ pairs. 
The production rates of $B^+B^-$ and $B^0\overline B^0$ pairs are assumed 
to be equal. 
The PID efficiency correction is determined 
using an inclusive $D^{*+}\to D^0\pi^+$, $D^0\to K^-\pi^+$ data sample 
where the track momenta and polar angles are required to be
consistent with those of $B^0\to \rho^0\rho^0$. 

The statistical significance is defined as 
$\mathcal{S}_0=\sqrt{-2\ln(\mathcal{L}_0 / \mathcal{L}_{\mathrm{max}})}$, 
where $\mathcal{L}_0$ and $\mathcal{L}_{\mathrm{max}}$ are the  
likelihoods of the fits with the signal yield fixed at zero and at the fitted 
value, respectively.

The 90\% confidence level (C.L.) upper limit is 
calculated from the equation
\begin{equation}
  {{\int_0^N \mathcal{L}(x) dx} \over
  {\int_0^{\infty}\mathcal{L}(x) dx} } = 90\% ,
\end{equation}
where $x$ indicates likelihood variables corresponding to the yield, and
$N$ is the upper bound for the yield that includes 90\% of the integral 
of the likelihood function.

The upper limit (UL) including systematic uncertainties is 
calculated by smearing the statistical likelihood function with a Gaussian,
where the Gaussian width is the combination of 
two total systematic errors: one is independent of the branching fraction and
the other is proportional to it. The significance including systematic 
uncertainties is calculated in the same way, but we only included the 
systematic errors related to signal yields in the convoluted Gaussian width.

\begin{figure}[htbp]
\centering
\epsfig{file=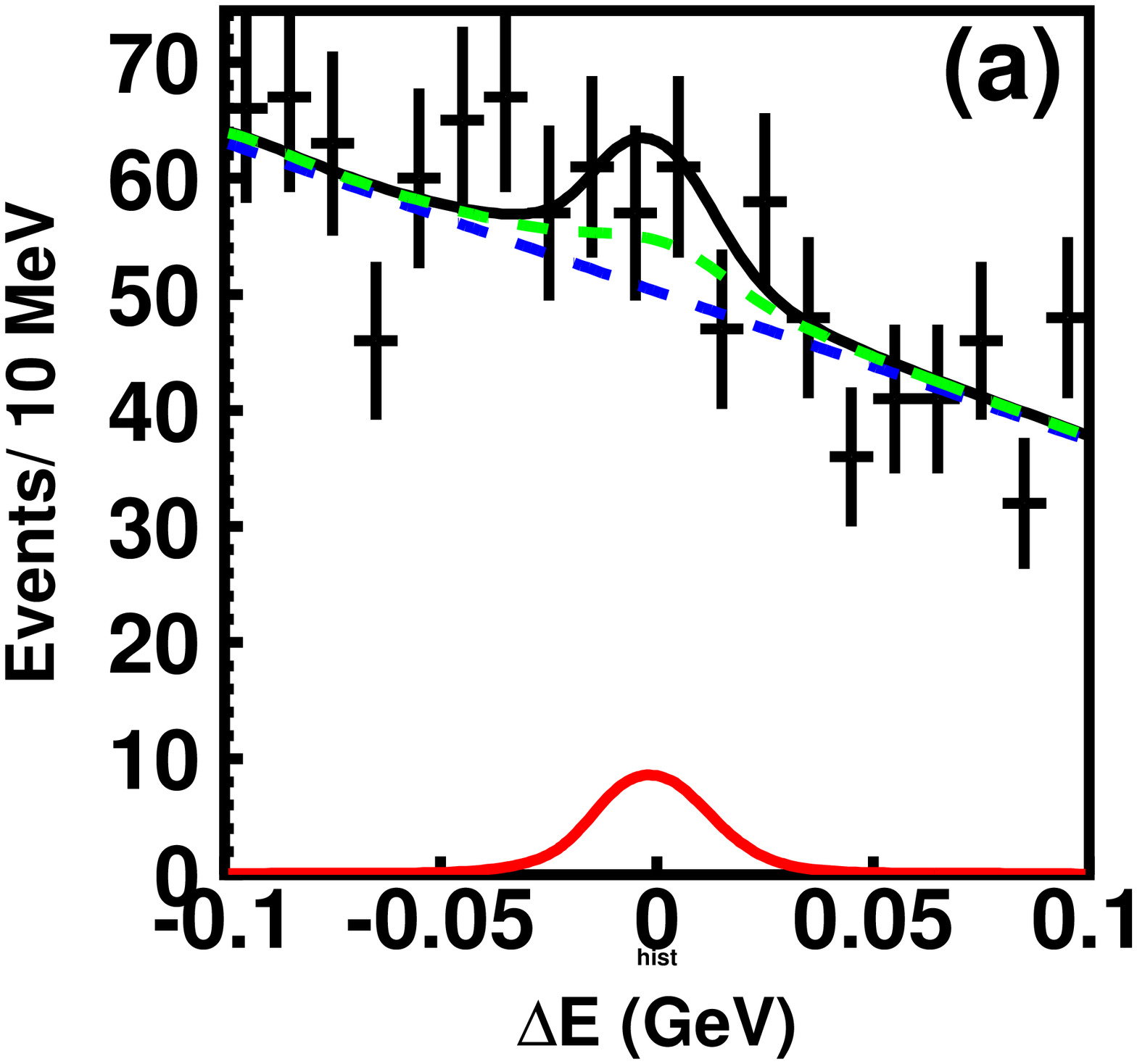,width=1.5in,height=1.5in}
\epsfig{file=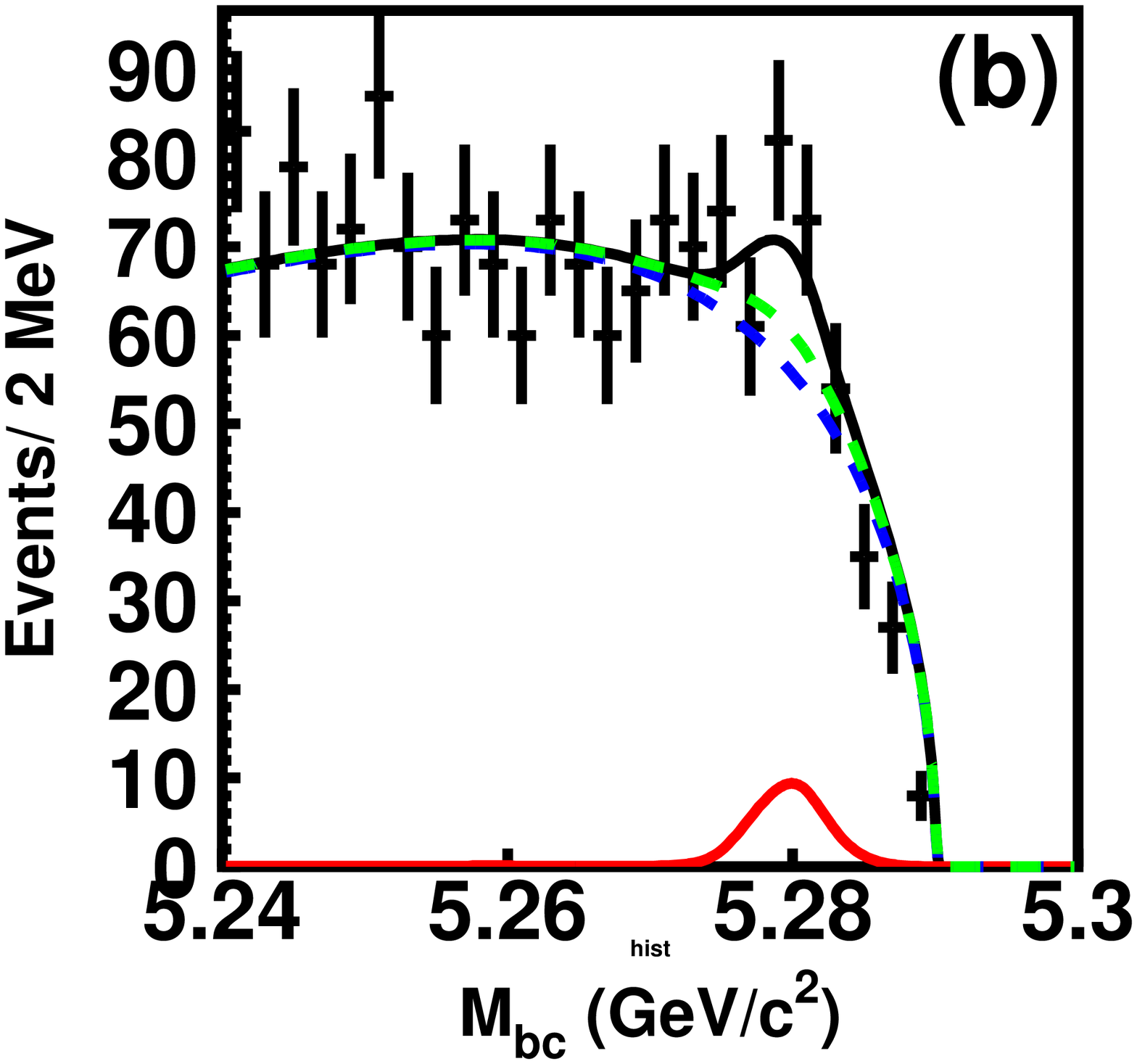,width=1.5in,height=1.5in}
\epsfig{file=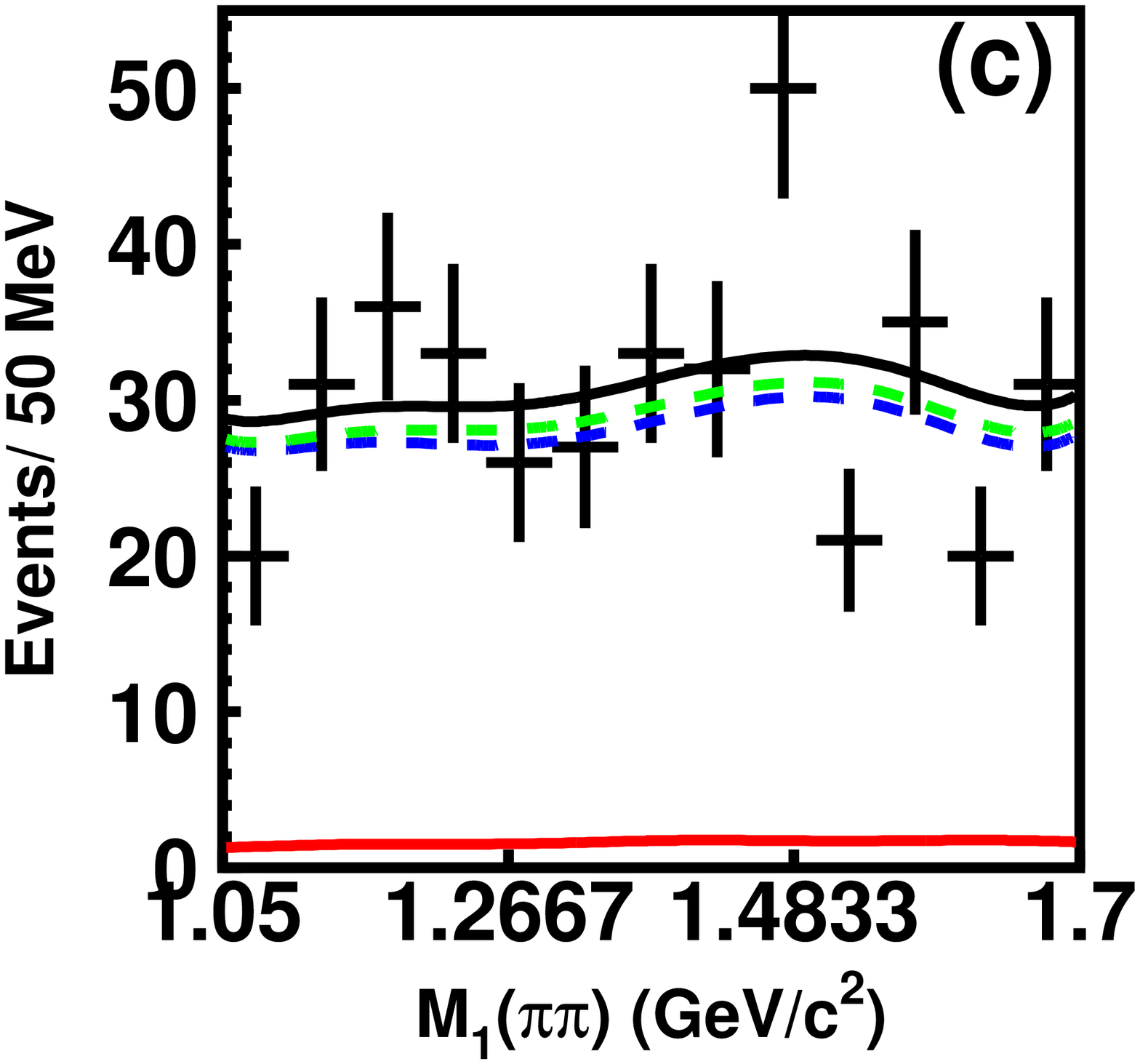,width=1.5in,height=1.5in}
\epsfig{file=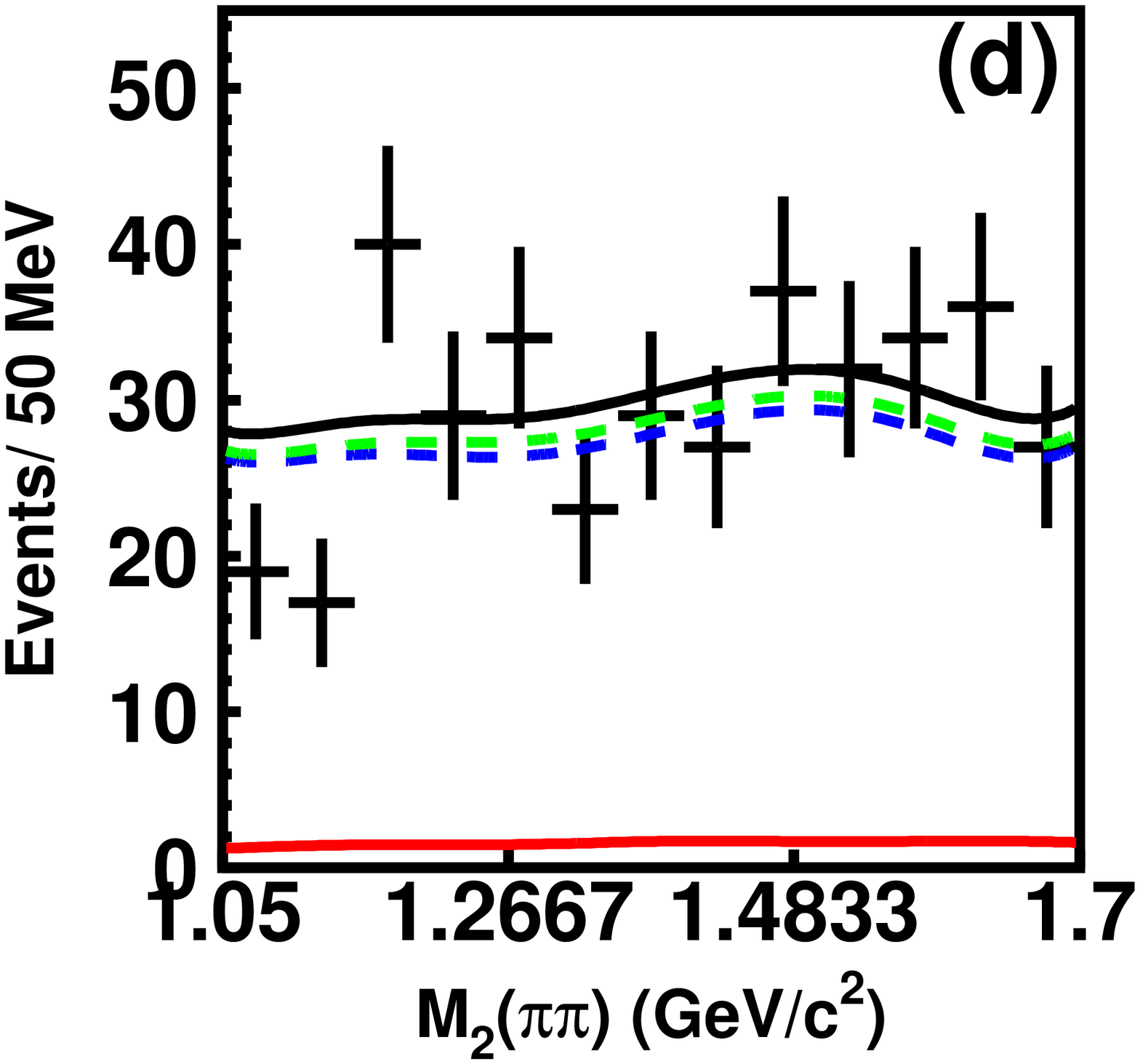,width=1.5in,height=1.5in}
\caption{Area A: projections of the four dimensional fit onto 
(a)$\Delta E$, (b)$M_{\mathrm{bc}}$, (c)$M_1(\pi\pi) \in$ (1.05, 1.7) 
GeV$/c^2$ and (d)$M_2(\pi\pi) \in$ (1.05, 1.7) GeV$/c^2$.
For the $\Delta E$ projection: 
$5.27\ \mathrm{GeV}/c^2 <M_{\mathrm{bc}}<5.29\ \mathrm{GeV}/c^2 $; 
for the $M_{\mathrm{bc}}$ projection:
$|\Delta E| < 0.05\ \mathrm{GeV}$; 
for the $M_{1(2)}(\pi\pi)$ projection:
$|\Delta E| < 0.05\ \mathrm{GeV}$ and 
$5.27\ \mathrm{GeV}/c^2 <M_{\mathrm{bc}}<5.29\ \mathrm{GeV}/c^2 $ and 
$1.35\ \mathrm{GeV}/c^2 <M_{2(1)}(\pi\pi)< 1.70\ \mathrm{GeV}/c^2$. 
The fit result is shown as the thick solid curve; the red solid 
curve represents the signal component, non-resonant $B^0\to 4\pi$ decay; 
the blue dashed and and green dashed curves represent the cumulative background 
components from continuum plus $b\to c$ backgrounds, and charmless $B$ decays.}
\label{fig-b2fourpi}
\end{figure}

\begin{figure}[htbp]
\centering
\epsfig{file=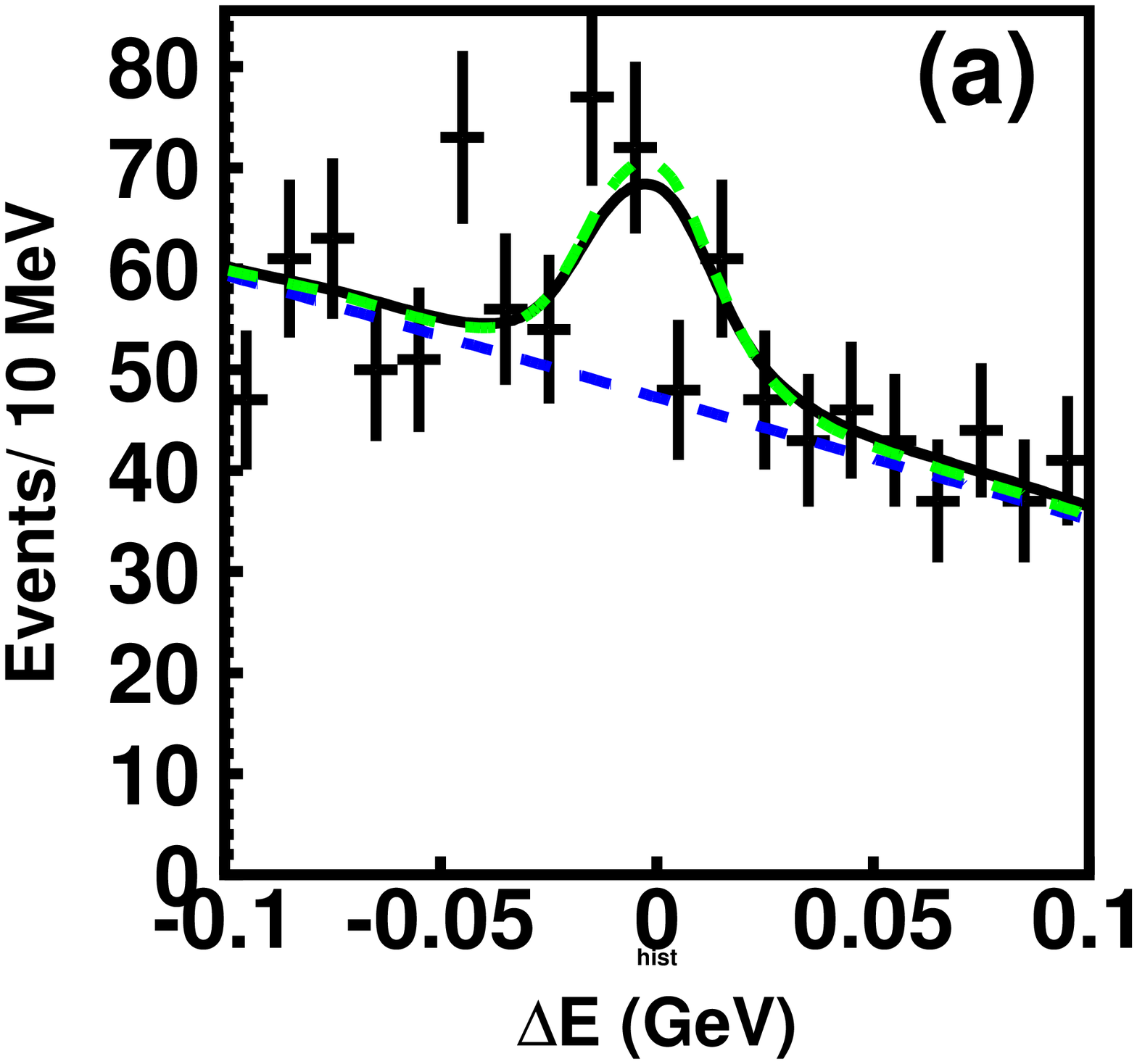,width=1.5in,height=1.5in}
\epsfig{file=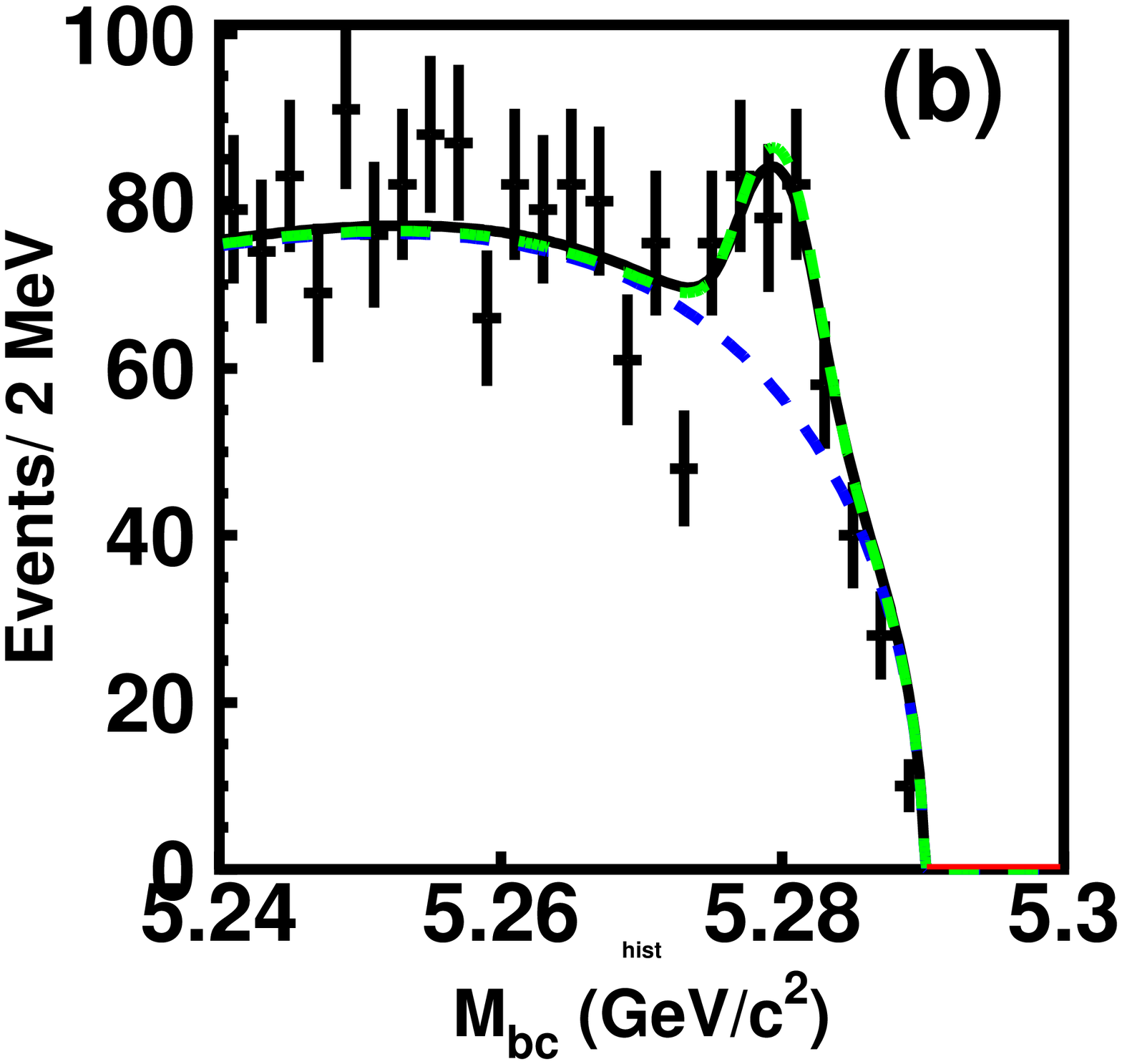,width=1.5in,height=1.5in}
\caption{Area B: projections of the two dimensional fit onto 
(a)$\Delta E$, (b)$M_{\mathrm{bc}}$. 
For the $\Delta E$ projection: 
$5.27\ \mathrm{GeV}/c^2 <M_{\mathrm{bc}}<5.29\ \mathrm{GeV}/c^2 $; 
for the $M_{\mathrm{bc}}$ projection:
$|\Delta E| < 0.05\ \mathrm{GeV}$. 
The fit result is shown as the thick solid curve; the $B^0\to \rho^0\pi\pi$ 
component cannot be seen here because of its negative yield; 
the blue dashed and and green dashed curves represent the cumulative background 
components from continuum plus $b\to c$ backgrounds, and charmless $B$ decays.}
\label{fig-b2rho0pipi}
\end{figure}

\begin{figure}[htbp]
\centering
\epsfig{file=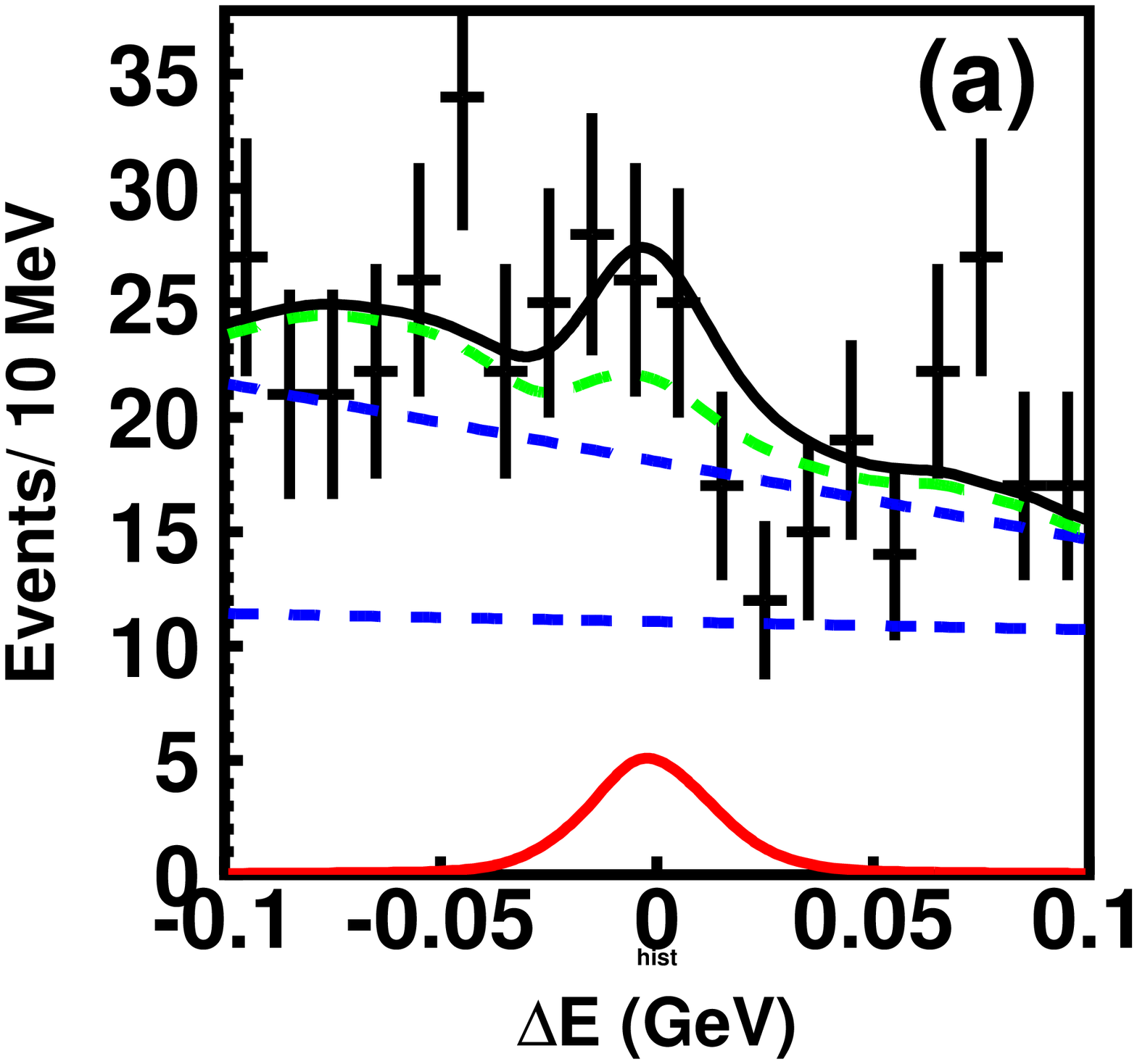,width=1.5in,height=1.5in}
\epsfig{file=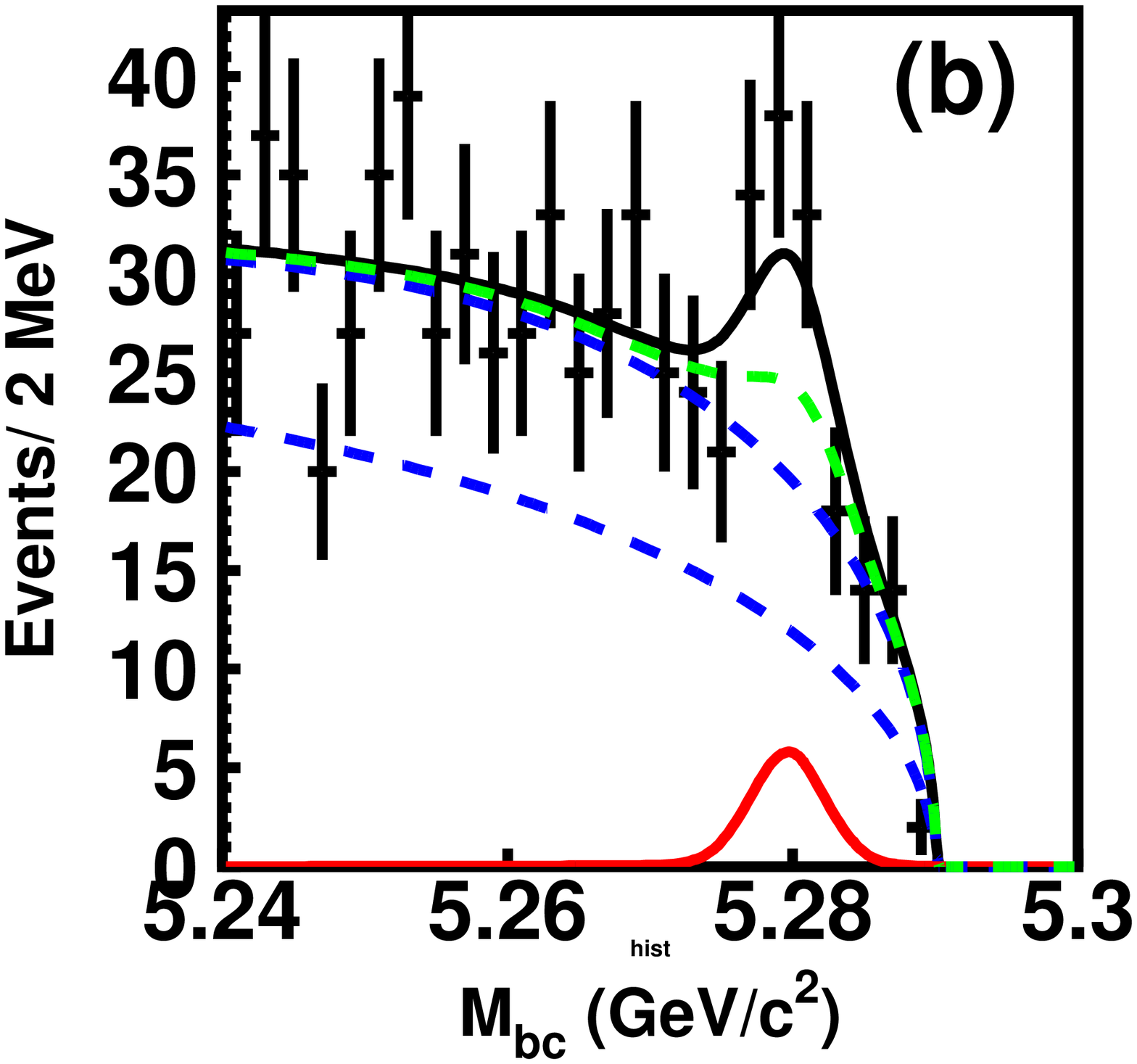,width=1.5in,height=1.5in}
\epsfig{file=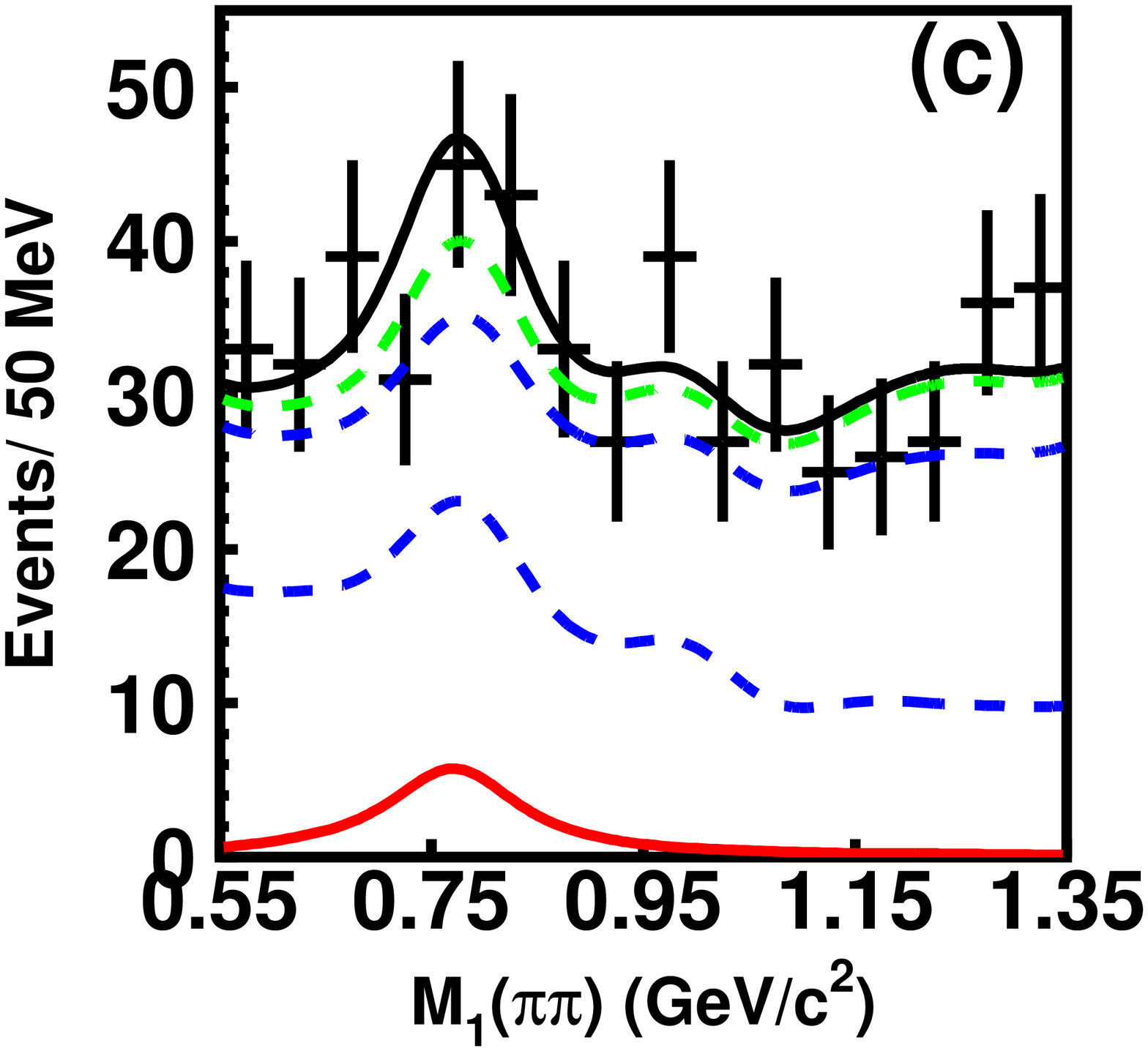,width=1.5in,height=1.5in}
\epsfig{file=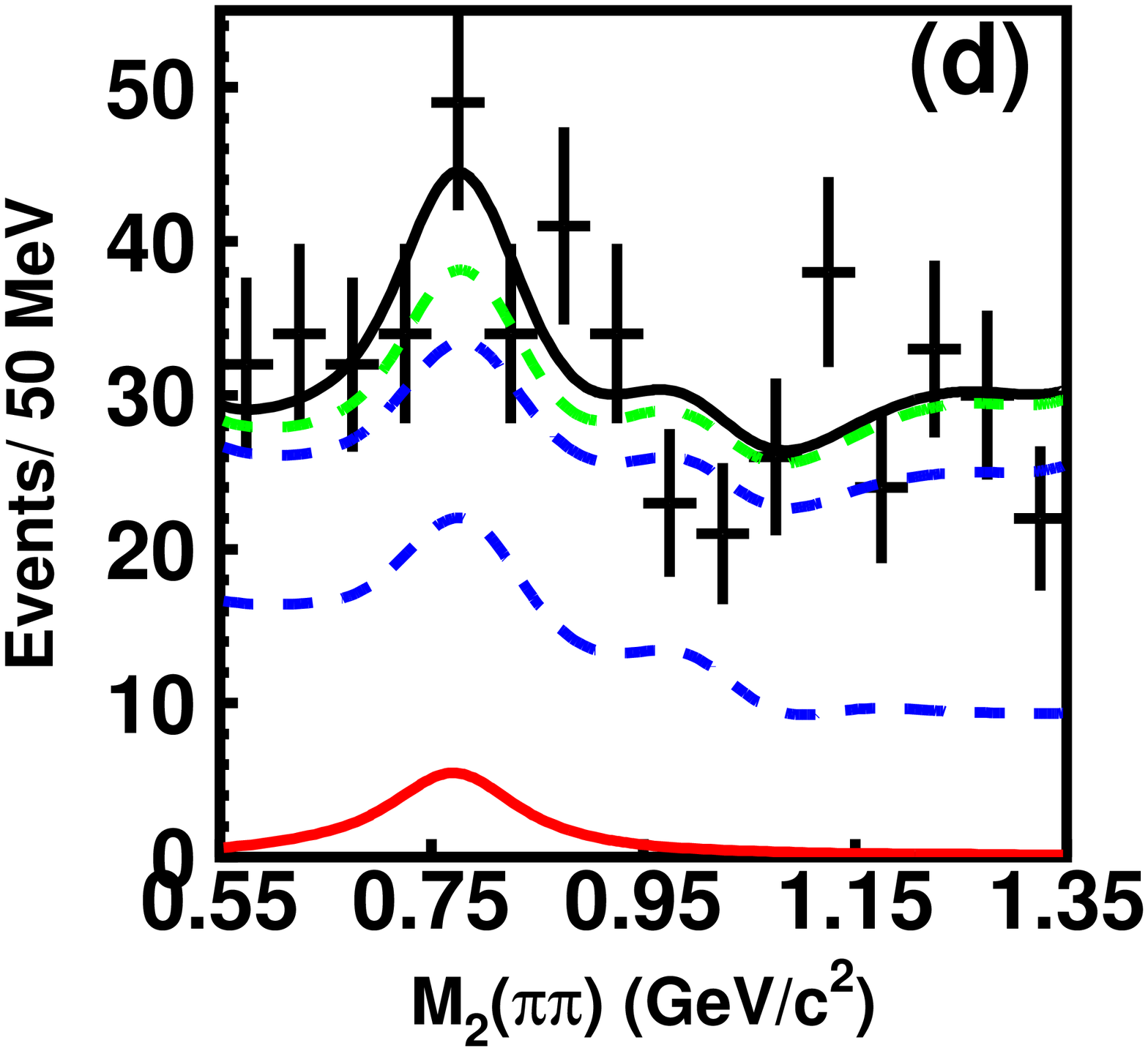,width=1.5in,height=1.5in}
\caption{Area C: projections of the four dimensional fit onto (a)$\Delta E$, 
(b)$M_{\mathrm{bc}}$, (c)$M_1(\pi\pi) \in$ (0.55, 1.35) GeV$/c^2$ and 
(d)$M_2(\pi\pi) \in$ (0.55, 1.35) GeV$/c^2$. 
For the $\Delta E$ projection: 
$5.27\ \mathrm{GeV}/c^2 <M_{\mathrm{bc}}<5.29\ \mathrm{GeV}/c^2 $; 
for the $M_{\mathrm{bc}}$ projection:
$|\Delta E| < 0.05\ \mathrm{GeV}$; 
for the $M_{1(2)}(\pi\pi)$ projection:
$|\Delta E| < 0.05\ \mathrm{GeV}$ and 
$5.27\ \mathrm{GeV}/c^2 <M_{\mathrm{bc}}<5.29\ \mathrm{GeV}/c^2 $ and 
$0.626\ \mathrm{GeV}/c^2 <M_{2(1)}(\pi\pi)< 0.926\ \mathrm{GeV}/c^2$. 
The fit result is shown as the thick solid curve; 
the red solid line represents the signal component, 
$B^0\to \rho^0\rho^0$ decay; the blue dashed, dash-dotted and green dashed 
curves represent, respectively, the cumulative background components from 
continuum processes, $b\to c$ decays, and charmless $B$ backgrounds.}
\label{fig-b2rho0rho0}  
\end{figure}

The fitted yields and branching fractions 
with systematic errors are listed in Table~\ref{table-yield}.
Fig.~\ref{fig-b2fourpi} shows the
projections of the data 
onto $\Delta E$, $M_{\mathrm{bc}}$, $M_1(\pi\pi)$ and $M_2(\pi\pi)$ for 
non-resonant $B^0\to 4\pi$ decay in area A. 
Fig.~\ref{fig-b2rho0pipi} shows the
projections of the data 
onto $\Delta E$ and $M_{\mathrm{bc}}$ for the $B^0\to \rho^0\pi\pi$ 
decay in area B. 
Fig.~\ref{fig-b2rho0rho0} shows the projections of the plots 
for the $B^0\to\rho^0\rho^0$ decay in area C. 
We measure $B^0 \to \rho^0\rho^0$ and 
non-resonant $B^0 \to 4\pi$ decays with 
1.7$\sigma$ and 2.2$\sigma$ significance, respectively. 
There are no significant yields for $B^0\to\rho^0f_0$, 
$B^0\to f_0f_0$, $B^0\to f_0\pi\pi$ or $B^0\to \rho^0\pi\pi$ decays.

\begin{table}[htbp]
\begin{center}
\caption{Fit results for each decay mode listed in the first column. 
The signal yields, reconstruction efficiency, 
significance including systematic uncertainties 
($\mathcal{S}$), branching fractions ($\mathcal{B}$) 
and UL including systematic uncertainties are listed. 
For the yields and branching fractions,
the first (second) error is statistical (systematic).}
\begin{tabular}{lcccccc} \hline \hline
Mode \ \       &  Yield                           & \ \ Eff.(\%)& \ \
$\mathcal{S}$ \ \ \ & $\mathcal{B}$($\times 10^{-6}$)  & UL($\times 10^{-6}$) (90\% C.L.) \cr \hline
\multicolumn{6}{c}{\bf Area A measurement:} \cr
$4\pi$         & $32.2\pm 14.9^{+7.1}_{-4.7}$     & 0.61        & 2.1 \                   & $10.2\pm4.7^{+2.3}_{-1.5}$       & $<$17.3 \cr \hline
\multicolumn{6}{c}{\bf Area B measurement:} \cr
$\rho^0\pi\pi$ & $-11.5\pm17.8^{+24.0}_{-15.0}$   & 0.86        & 0.0 \                   & $-$                              & $<$6.3  \cr \hline
\multicolumn{6}{c}{\bf Area C measurement:} \cr
$\rho^0\rho^0$ & $33.7\pm16.0^{+12.5}_{-13.2}$    & 7.11        & 1.8 \                   & $0.9\pm0.4^{+0.3}_{-0.4}$     & $<$1.6  \cr
$\rho^0f_0$    & $3.6\pm11.7^{+7.3}_{-7.7}$       & 4.01        & 0.3 \                   & $0.2\pm0.6\pm0.4$                & $<$1.0  \cr
$f_0f_0$       & $1.6\pm3.7\pm2.9$                & 2.23        & 0.4 \                   & $0.1\pm0.3\pm0.3$                & $<$0.8  \cr
$f_0\pi\pi$    & $-6.0\pm19.7^{+23.0}_{-16.6}$    & 0.71        & 0.0 \                   & $-$                              & $<$8.6  \cr
\hline \hline
\end{tabular}
\label{table-yield}
\end{center}
\end{table}

\begin{center}
{\bf $\rho^0$ HELICITY \\}
\end{center}

\begin{figure}[htbp]
\centering
\epsfig{file=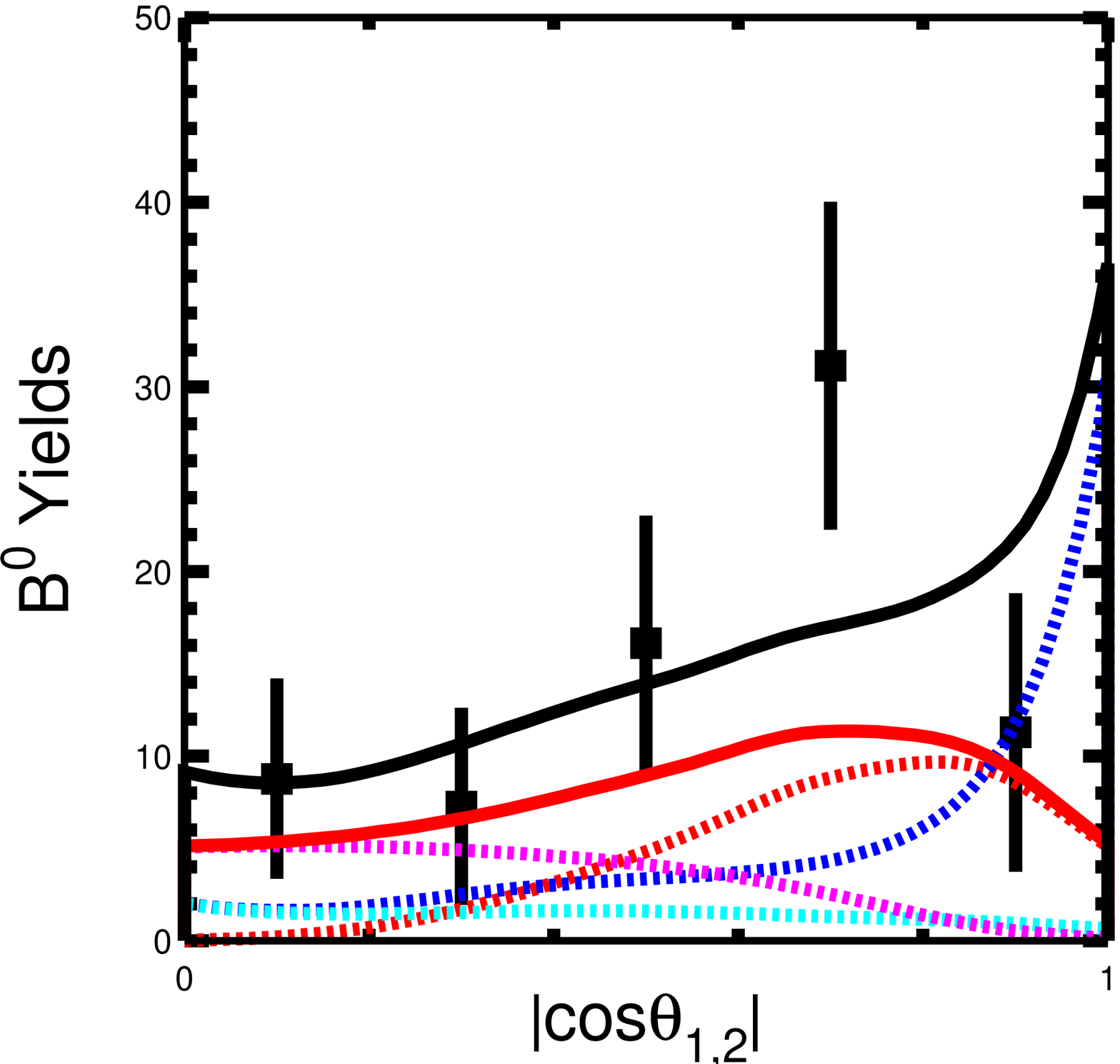,width=2.in,height=2.1in}
\caption{Sum of two $\rho^0$ helicity angle distributions
for background-subtracted events in the region
($0.626\ \mathrm{GeV}/c^2<M_{1,2}(\pi\pi)<0.926\ \mathrm{GeV}/c^2$). 
The points with error bars represent the $B^0$ 
yields obtained from $\Delta E$-$M_{\mathrm{bc}}$ ML fits 
for each helicity angle interval.
The black overlaid curve is the sum of expected distributions 
from MC. 
The blue, cyan, red and magenta dashed curves represent the $B^0\to a_1\pi$, 
non-resonant $B^0\to 4\pi$, longitudinally and transversely polarized 
$B^0\to\rho^0\rho^0$ decays, respectively.
The red solid curve represents the sum of two polarized $B^0\to\rho^0\rho^0$ 
decays.}
\label{fig-rho0rho0hel}
\end{figure}

We perform an angular analysis of $B^0\to\rho^0\rho^0$ using 
the sum of two $\rho^0$ helicity angle distributions 
for the signal candidates having
$M_1(\pi\pi)$ and $M_2(\pi\pi)$ values in the signal region 
($0.626\ \mathrm{GeV}/c^2<M_{1,2}(\pi\pi)<0.926\ \mathrm{GeV}/c^2$).
The $\rho^0$ helicity angle is defined as the angle between 
the $\pi^+$ direction and the 
$B^0$ direction in the $\rho^0$ rest frame. 
Fig.~\ref{fig-rho0rho0hel}
shows the sum of the two $\rho^0$ helicity angle distributions;
each bin has a $B^0$ yield obtained from 
a $\Delta E$-$M_{\mathrm{bc}}$ fit that does not
distinguish whether the $B^0$ decays into
$\rho^0\rho^0$, $a_1^{\pm}\pi^{\mp}$ or $4\pi$.
We fix the fractions of the three $B^0$ decays modes
to values obtained from the the 4-D ML fit in area C.
In the helicity angle fit,
we vary the longitudinal 
polarization fraction for the $B^0\to\rho^0\rho^0$ component. 
A $\chi^2$ fit yields a longitudinal polarization fraction for 
$B^0\to\rho^0\rho^0$ of $f_L=0.6\pm0.2$.
The statistical error is obtained from a MC pseudo-experiment study,
since the two $\rho^0$ helicity angle distributions are correlated.


\begin{center}
\begin{Large}
{\bf SYSTEMATIC ERROR \\}
\end{Large}
\end{center}

The main systematic uncertainty for the branching fraction of
non-resonant $B^0\to 4\pi$ decay is the uncertainty of 
the $B^0\to a_1^{\pm}\pi^{\mp}$ branching fraction.
The main systematic uncertainties for the $B^0\to\rho^0\pi\pi$ branching 
fraction are the uncertainties of the non-resonant $B^0\to 4\pi$ and 
$B^0\to a_1^{\pm}\pi^{\mp}$ branching fractions.
For $B^0\to\rho^0\rho^0$ decay, 
the main sources of systematic uncertainties 
include the uncertainties on the branching fractions of
$B^0\to a_1^{\pm}\pi^{\mp}$, non-resonant $B^0\to 4\pi$, $B^0\to\rho^0\pi\pi$, 
$B^{\pm}\to \rho^0\rho^{\pm}$ and $B^0\to \rho^0 K^{*0}$. 

We vary the branching fractions of $B^0\to a_1^{\pm}\pi^{\mp}$ 
($33.2\pm 4.8$, in units of $10^{-6}$)~\cite{107} and 
$B^{\pm}\to \rho^0\rho^{\pm}$ ($18.2\pm3.0$)~\cite{108}
by their $\pm1\sigma$ errors. 
On the other hand, we vary the branching fractions of 
$B^{0}\to \rho^0K^{*0}$, non-resonant $B^0\to 4\pi$ and $B^0\to \rho^0\pi\pi$ 
in the range $(0, 7.2)\times 10^{-6}$~\cite{108}, $(0, 15.4)\times 10^{-6}$ 
(area A measurement) and $(0, 4.1)\times 10^{-6}$ 
(area B measurement), respectively, 
because these branching fractions are not measured with high significance. 
The fits are repeated and the differences between 
the results and the nominal fit values are taken as systematic errors. 

According to MC, the signal SCF fractions are 
16.2\% for $B^0\to\rho^0\rho^0$, 10.6\% for $B^0\to\rho^0f_0$, 7.3\% for 
$B^0\to f_0f_0$, 12.4\% for $B^0\to f_0\pi\pi$, 10.8\% for 
$B^0\to\rho^0\pi\pi$ and 11.1\% for non-resonant $B^0\to 4\pi$. 
We estimate a systematic uncertainty for the signal SCF 
by setting its fraction to zero. 
A systematic error for the longitudinal polarization fraction of 
$B^0\to\rho^0\rho^0$ is obtained by changing the fraction from the nominal 
value $f_L=1$ to our measured value $f_L=0.6$.
A MC study indicates that the fit biases are 
$+7$ events for $B^0\to\rho^0\rho^0$, 
$+1$ event for $B^0\to\rho^0f_0$, $-1$ event for $B^0\to f_0f_0$, 
$-3$ events for $B^0\to f_0\pi\pi$ and $+6$ events for 
non-resonant $B^0\to 4\pi$.
We find that fit biases occur due to the correlations between the two sets 
of variables ($\Delta E$, $M_{\rm bc}$) and ($M_1$ and $M_2$), 
which are not taken into account in our fit. 
We correct the yields in the fit for these biases and include the 
corrections as systematic errors.

Systematic uncertainties for the $\Delta E$-$M_{\rm bc}$ PDFs used in the fit 
are estimated by performing the fits while varying the 
signal peak positions and resolutions by $\pm 1\sigma$. 
Systematic uncertainties for the $M_1$-$M_2$ PDFs are estimated in a similar 
way. We vary the mean and width of the $\rho^0$ and $f_0$ masses in the 
$M_1$-$M_2$ PDFs for the decay modes $B^0\to\rho^0\rho^0$, $B^0\to\rho^0f_0$, 
$B^0\to f_0f_0$, and $B^0\to f_0\pi\pi$. For the non-resonant $B^0\to 4\pi$, 
other rare $B$ decay, $b \to c$ and continuum backgrounds, we vary their 
$M_1$-$M_2$ PDF shapes, repeat the fits, and take the deviations from the 
central values as the systematic errors.

We test the possible interference between $B^0\to a_1^{\pm}\pi^{\mp}$, 
non-resonant $B^0\to 4\pi$ and $B^0\to\rho^0\rho^0$ by toy MC. 
We add a simple interference model to the toy MC generation, 
which is, for $\rho^0\to\pi^+\pi^-$ decay, modified from a relativistic 
Breit-Wigner function to
\begin{equation}
\small
   \Biggl\vert {1 \over {m^2-m_0^2+ im_0\Gamma}} 
   + A\mathrm{e}^{-i\delta}\Biggr\vert^2 
   = A^2 + 2A\Biggl[ {{(m^2-m_0^2)\cos\delta - \Gamma m_0\sin\delta} 
     \over {(m^2-m_0^2)^2 + (\Gamma m_0)^2 }} \Biggr]
     + {1 \over {(m^2-m_0^2)^2+(\Gamma m_0)^2}} ,
\end{equation}
where $A$ and $\delta$ are the interference amplitude and phase, $m_0$ and
$\Gamma$ are the $\rho^0$ mass and width, respectively.
We assume that the interference term due to the amplitudes for 
$B^0\to a_1^{\pm}\pi^{\mp}$ and non-resonant $B^0\to 4\pi$ decays is constant in
the $B^0\to\rho^0\rho^0$ signal region. Since the magnitude of the interfering
amplitude and relative phase are not known, we uniformly vary these parameters
and perform a fit in each case to measure the deviations from the incoherent
case. The mean deviation is calculated, and we add and subtract the r.m.s. of 
the distribution of deviations from this value to obtain the systematic
uncertainty. The interference systematic for $B^0\to \rho^0\rho^0$ decay 
is $(^{+12.6}_{-4.8})$\%.

The systematic errors for the efficiency
arise from the tracking efficiency, particle identification (PID) and  
$\mathcal{R}$ requirement. 
The systematic error due to the track finding efficiency is 
estimated to be 1.3\% per track using partially
reconstructed $D^*$ events. 
The systematic error due to the pion identification (PID) is 
1.2\% per track estimated using an inclusive $D^*$ control sample. 
The $\mathcal{R}$ requirement systematic error is determined 
from the efficiency difference between data and MC 
using a $B^0\to D^+(K\pi\pi)\pi^-$ control sample. 
Table II summarizes the sources of systematic uncertainties
and their quadratic sum for each of the items. 
The overall relative systematic errors are $(^{+37.1}_{-39.1})$\% for
the $B^0\to \rho^0\rho^0$ decay and 
$(^{+22.2}_{-14.5})$\% for non-resonant $B^0\to 4\pi$ decay.
\begin{table}[htbp]
\begin{center}
\caption{Summary of systematic errors (\%) for 
the branching fraction measurements. $f_L$ and $f_{SCF}$ 
are the fractional uncertainties for longitudinal polarization and 
self-cross-feed.}
\vspace*{0.2cm}
\begin{tabular}{lcccccccccc} \hline \hline
Source                                     & 4$\pi$             & $\rho^0\pi\pi$       & $\rho^0\rho^0$     & $\rho^0f_0$         &  $f_0f_0$          & $f_0\pi\pi$          \cr \hline
Fitting PDF                                & $\pm$6.8           & $\pm$9.6             & $\pm$13.1          & $\pm$197.7          &  $\pm$159.4        & $\pm$194.4           \cr
$\mathcal{B}(B^0\to a_1\pi)$               & $\pm$3.7           & $^{+100.0}_{-96.5}$  & $^{+4.7}_{-5.3}$   & $^{+2.8}_{-5.6}$    &  $^{+18.8}_{-12.5}$& $^{+48.3}_{-51.7}$   \cr
$\mathcal{B}(B^0\to 4\pi)$                 & $-$                & $^{+181.7}_{-86.1}$  & $^{+14.5}_{-7.1}$  & $^{+16.7}_{-52.8}$  &  $^{+37.5}_{-50.0}$& $^{+326.7}_{-146.7}$ \cr
$\mathcal{B}(B^0\to \rho^0\pi\pi)$         & $-$                & $-$                  & $-28.5$            & $-55.6$             &  $+31.3$           & $-108.3$             \cr
$\mathcal{B}(B^{\pm}\to \rho^0\rho^{\pm})$ & $-$                & $0.0$                & $\pm$0.6           & $\pm$2.8            &  $0.0$             & $0.0$                \cr
$\mathcal{B}(B^0\to \rho^0 K^{*0})$        & $-$                & $-$                  & $+1.5$             & $+13.9$             &  0.0               & $-16.7$              \cr
$f_L$                                      & $-$                & $-$                  & $-11.1$            & $ -$                &  $-$               & $-$                  \cr
$f_{SCF}$                                  & $-9.6$             & $+13.9$              & $-15.7$            & $-11.1$             &  $+18.8$           & $+3.3$               \cr
Interference                               & $-$                & $-$                  & $^{+12.6}_{-4.8}$  & $-$                 & $-$                & $-$                  \cr          
Fit bias correction                        & $^{+19.6}_{-3.1}$  & $\pm8.7$             & $^{+22.0}_{-3.0}$  & $^{+36.0}_{-27.8}$  &  $\pm$62.5         & $^{+16.7}_{-48.3}$   \cr
Tracking                                   & $\pm$4.2           & $\pm$4.3             & $\pm$5.1           & $\pm$5.0            &  $\pm$4.4          & $\pm$4.3             \cr
PID                                        & $\pm$4.1           & $\pm$4.2             & $\pm$4.9           & $\pm$4.6            &  $\pm$4.5          & $\pm$4.2             \cr
$\mathcal{R}$ requirement                  & $\pm$3.4           & $\pm$3.4             & $\pm$3.4           & $\pm$3.4            &  $\pm$3.4          & $\pm$3.4             \cr
$N_{B \overline B}$                        & $\pm$1.3           & $\pm$1.3             & $\pm$1.3           & $\pm$1.3            &  $\pm$1.3          & $\pm$1.3             \cr \hline
Sum(\%)                                    & $^{+22.2}_{-14.5}$ & $^{+208.4}_{-130.2}$ & $^{+37.1}_{-39.1}$ & $^{+202.3}_{-214.4}$&$^{+180.2}_{-179.0}$& $^{+383.7}_{-276.4}$ \cr
\hline \hline
\end{tabular}
\end{center}
\end{table}

\begin{center}
\begin{Large}
{\bf SUMMARY \\}
\end{Large}
\end{center}

In summary, we measure the branching fraction of $B^0\to\rho^0\rho^0$ to 
be $(0.9\pm0.4^{+0.3}_{-0.4})\times 10^{-6}$ with 1.8$\sigma$
significance; the 90\% confidence level upper 
limit including systematic uncertainties is 
$\mathcal{B}(B^0\to\rho^0\rho^0)<1.6 \times 10^{-6}$. 
Since no significant signal is found, we have assumed this mode is a 
longitudinally polarized decay ($f_L=1$), to obtain the most conservative 
upper limit. 
For $f_L=0.0$, we obtain 
the central value of $\mathcal{B}(B^0\to\rho^0\rho^0)=0.6 \times 10^{-6}$.
Measurements of polarization and asymmetry in $B^0\to\rho^0 \rho^0$ will be 
needed to improve the precision of the $\phi_2$ constraint. 

On other hand, we find an excess in non-resonant 
$B^0\to\ 4\pi$ decay with 2.1$\sigma$ significance. 
We measure the branching fraction and a 90\% confidence level upper limit 
for non-resonant $B^0\to\ 4\pi$ decay to be 
$(10.2\pm4.7^{+2.3}_{-1.5})\times 10^{-6}$ and 
$\mathcal{B}(B^0\to 4\pi)<17.3 \times 10^{-6}$. 
This contribution was not taken into 
account in previous measurements of $B^0\to\rho^0\rho^0$ \cite{11}. 
We find no significant signal for the decays 
$B^0\to\rho^0f_0$, $B^0\to\ f_0f_0$, $B^0\to\ f_0\pi\pi$ and 
$B^0\to \rho^0\pi\pi$; the corresponding upper limits are 
listed in Table I.

\begin{center}
\begin{Large}
{\bf ACKNOWLEDGMENTS \\}
\end{Large}
\end{center}

We thank the KEKB group for the excellent operation of the
accelerator, the KEK cryogenics group for the efficient
operation of the solenoid, and the KEK computer group and
the National Institute of Informatics for valuable computing
and Super-SINET network support. We acknowledge support from
the Ministry of Education, Culture, Sports, Science, and
Technology of Japan and the Japan Society for the Promotion
of Science; the Australian Research Council and the
Australian Department of Education, Science and Training;
the National Science Foundation of China and the Knowledge
Innovation Program of the Chinese Academy of Sciences under
contract No. 10575109 and IHEP-U-503; the Department of
Science and Technology of India; 
the BK21 program of the Ministry of Education of Korea, 
the CHEP SRC program and Basic Research program 
(grant No. R01-2005-000-10089-0) of the Korea Science and
Engineering Foundation, and the Pure Basic Research Group 
program of the Korea Research Foundation; 
the Polish State Committee for Scientific Research; 
the Ministry of Education and Science of the Russian
Federation and the Russian Federal Agency for Atomic Energy;
the Slovenian Research Agency; the Swiss
National Science Foundation; the National Science Council
and the Ministry of Education of Taiwan; and the U.S. 
Department of Energy.



\end{document}